\documentclass[aps,pra,floatfix,twocolumn,showpacs,10pt]{revtex4-1}

\usepackage{graphicx}
\usepackage{dcolumn}
\usepackage{bm}
\usepackage{amssymb}
\usepackage{rotating} 
\usepackage{xcolor}

\begin{document}


\title{Dual-Cone Variational Calculation of the 2-Electron Reduced Density Matrix}

\author{David A. Mazziotti}

\email{damazz@uchicago.edu}
\affiliation{Department of Chemistry and The James Franck Institute, The University of Chicago, Chicago, IL 60637}%

\date{Submitted July 27, 2020\textcolor{black}{; Revised October 13, 2020}}


\begin{abstract}

The computation of strongly correlated quantum systems is challenging because of its potentially exponential scaling in the number of electron configurations.  Variational calculation of the two-electron reduced density matrix (2-RDM) without the many-electron wave function exploits the pairwise nature of the electronic Coulomb interaction to compute a lower bound on the ground-state energy with polynomial computational scaling.  Recently, a dual-cone formulation of the variational 2-RDM calculation was shown to generate the ground-state energy, albeit not the 2-RDM, at a substantially reduced computational cost, especially for higher $N$-representability conditions such as the T2 constraint.  Here we generalize the dual-cone variational 2-RDM method to compute not only the ground-state energy but also the 2-RDM.  The central result is that we can  compute the 2-RDM from a generalization of the Hellmann-Feynman theorem.  Specifically, we prove that in the Lagrangian formulation of the dual-cone optimization the 2-RDM is the Lagrange multiplier.   We apply the method to computing the energies and properties of strongly correlated electrons---including atomic charges, electron densities, dipole moments, and orbital occupations---in an illustrative hydrogen chain and the nitrogen-fixation catalyst FeMoco.  The dual variational computation of the 2-RDM with T2 or higher $N$-representability conditions provides a polynomially scaling approach to strongly correlated molecules and materials with significant applications in  atomic and molecular and condensed-matter chemistry and physics.

\end{abstract}

\pacs{31.10.+z}

\maketitle

\section{Introduction}

Strong electron correlation can be critically important in the accurate prediction of energies and properties of molecules and materials including conjugated organic molecules, transition-metal catalysts, spintronic molecules, and superconductors.  Strongly correlated quantum systems arise when a linear increase in system size produces an exponentially increasing number of electron configurations that contribute significantly to the quantum-mechanical wave function~\cite{Mazziotti2007,Coleman2000,Huang2006,Horodecki2009}.  Traditional wave function methods that are based upon a single-reference determinant such as density functional theory~\cite{Cohen2011} and coupled cluster theory~\cite{Bartlett2007} can have difficulty in describing strongly correlated wave functions.  Recent advances in the description of such wave functions include density matrix renormalization group~\cite{Schollwoeck2005} as well as sparse configuration interaction methods~\cite{Booth2009,Li2018,Wang2019}.  An alternative approach to strong correlation is the direct variational calculation of the two-electron reduced density matrix (2-RDM) without the computation or storage of the many-electron wave function~\cite{Mazziotti2007, Coleman2000, Erdahl2000, Nakata2001, Mazziotti2001c, Mazziotti2002b, Zhao2004, Mazziotti2004a, Cances2006, Fukuda2007, Gidofalvi2008, Verstichel2009, Shenvi2010, Mazziotti2011, Verstichel2012, Baumgratz2012, Schilling2013, Veeraraghavan2014a, Poelmans2015, Fosso-Tande2016, Mazziotti2016, Piris2017a, Alcoba2018, Rubio-Garcia2019, Benavides-Riveros2020, Head-Marsden2020, Haim2020, Han2020}.  The 2-RDM method exploits the pairwise nature of the electron-electron interaction in the electronic Hamiltonian. Variational calculation of the 2-RDM has been applied to the accurate computation of a range of strongly correlated phenomena including polyradical character in conjugated polyaromatic hydrocarbons~\cite{Gidofalvi2008}, non-innocent ligand effects in transition-metal complexes\cite{Schlimgen2016, McIsaac2017}, entanglement-driven non-superexchange mechanisms in bridged transition-metal dimers~\cite{Boyn2020, Xie2020}, and exciton condensation in molecular-scale electron double layers~\cite{Safaei2018, Sager2020a}.

The 2-RDM must be constrained by conditions to ensure that it represents at least one $N$-electron density matrix known as $N$-representability conditions\cite{Mazziotti2012b, Coleman1963, Garrod1964, Kummer1967, Erdahl1978}.  Most applications of the variational 2-RDM method employ a set of $N$-representability constraints known as the two-positivity conditions.  The two-positivity conditions are part of a hierarchy of $p$-positivity conditions in which $p$-body metric matrices are constrained to be positive semidefinite~\cite{Mazziotti2001c, Mazziotti2012b}.  While the two-positivity conditions typically generate an accurate lower bound to the ground-state energy, the bound can often be significantly improved through three-positivity or partial three-positivity conditions such as the T2 condition~\cite{Erdahl2000,Mazziotti2001c,Zhao2004,Mazziotti2005,Mazziotti2006b,Verstichel2012}.  In the conventional (primal) formulation of the variational 2-RDM method in which $N$-representability constraints are placed directly on the 2-RDM, however, the computational cost $r^{9}$ of the three-positivity conditions including T2 is much greater than the cost $r^{6}$ of the 2-positivity conditions~\cite{Mazziotti2005,Mazziotti2006b}.  Recently, we proposed and implemented a dual formulation of the variational 2-RDM method with an $r^6$ scaling for the T2 condition~\cite{Mazziotti2016} in which the lower bound to the ground-state energy is directly computed by fitting the $N$-representability conditions to the Hamiltonian.  In its original formulation~\cite{Mazziotti2016}, however, this dual approach generates only the ground-state energy.  In this paper we show theoretically and computationally how the dual-cone approach can be extended to compute not only the energy but also the 2-RDM.

To obtain the 2-RDM in the dual-cone approach, we employ an extension of the Hellmann-Feynman theorem for 2-RDM theory.  While the proof of the Hellmann-Feynman theorem is well-known for wave functions~\cite{Feynman1939}, it must be generalized for 2-RDM theory to treat not only $N$-representable 2-RDMs but also approximately $N$-representable 2-RDMs~\cite{Schlimgen2018}.  With this extension we show that the derivative of the energy from the variational 2-RDM method with respect to the reduced Hamiltonian matrix generates the 2-RDM.  Using this relation, we prove the central result of the paper that in the Lagrangian of the dual-cone optimization the 2-RDM is the Lagrange multiplier.  This result allows us to compute the 1- and 2-RDMs as well as all one- and two-body properties efficiently in a dual formulation of the variational 2-RDM theory.   To illustrate method, we apply the dual 2-RDM method with the T2 condition to computing the Mott metal-insulator transition of the hydrogen chain (H$_{4}$) as well as the electronic structure of the strongly correlated, nitrogen-fixation catalyst FeMoco.

\section{Theory}

After a review of the primal formulation of variational 2-RDM theory in section~\ref{sec:primal}, we present a generalization of the Hellmann-Feynman theorem for 2-RDM theory in section~\ref{sec:hf}, which we use in section~\ref{sec:dual} to derive a relation for the 2-RDM in the dual formulation of variational 2-RDM theory.

\subsection{Primal Formulation of 2-RDM Theory}

\label{sec:primal}

For a many-particle quantum system with only pairwise interactions the minimization of the ground-state energy as a functional of the 2-RDM can be written as~\cite{Mazziotti2007, Coleman2000, Erdahl2000, Nakata2001, Mazziotti2001c, Mazziotti2002b, Zhao2004, Mazziotti2004a, Cances2006, Fukuda2007, Gidofalvi2008, Verstichel2009, Shenvi2010, Mazziotti2011, Verstichel2012, Baumgratz2012, Schilling2013, Veeraraghavan2014a, Poelmans2015, Fosso-Tande2016, Mazziotti2016, Piris2017a, Alcoba2018, Rubio-Garcia2019, Benavides-Riveros2020, Head-Marsden2020, Coleman1963, Garrod1964, Kummer1967, Erdahl1978}
\begin{equation}
\label{eq:minE}
E^{*} = \min_{^{2} D \in {\tilde P}^{2}_{N}}{ {\rm Tr}{ \left (^{2} K \, ^{2} D \right ) } }
\end{equation}
where $E^{*}$ is the energy at the minimum for a given two-electron reduced Hamiltonian matrix $^{2} K$.  For a finite basis of $r$ orbitals the two-particle reduced Hamiltonian and density matrices are square Hermitian matrices of dimension $r(r-1)/2$.  Minimization is performed with respect to an approximate set ${\tilde P}^{2}_{N}$ of ensemble $N$-representable 2-RDMs.   We define ${\tilde P}^{2}_{N}$ to be convex and a superset of the exact convex set $P^{2}_{N}$ of ensemble $N$-representable 2-RDMs, that is $P^{2}_{N} \subseteq  {\tilde P}^{2}_{N}$.   A 2-RDM is ensemble $N$-representable if and only if it is representable by at least one $N$-particle density matrix~\cite{Coleman2000, Mazziotti2012b, Garrod1964, Kummer1967, Erdahl1978}.   Because $P^{2}_{N} \subseteq  {\tilde P}^{2}_{N}$, the minimum energy $E^{*}$ is a lower bound to the exact ground-state energy of the Schr{\"o}dinger equation in the finite basis set with the $N$-particle Hamiltonian corresponding to $^{2} K$.  As the approximate set ${\tilde P}^{2}_{N}$ approaches the exact set $P^{2}_{N}$, the minimum energy $E^{*}$ approaches the exact ground-state energy from below.

\subsection{Hellmann-Feynman Theorem of the 2-RDM}

\label{sec:hf}

We can derive an extension of the Hellmann-Feynman theorem for approximate $N$-representable sets of 2-RDMs.  Consider the derivative of the minimum energy in Eq.~(\ref{eq:minE}) with respect to an arbitrary parameter $R$ to obtain
\begin{eqnarray}
\frac{\partial E^{*}}{\partial R} & = & {\rm Tr}{\left ( \frac{\partial \left (^{2} K \right )}{\partial R} \, ^{2} D^{*}  \right )} + {\rm Tr}{ \left ( ^{2} K \, \frac{\partial \left ( ^{2} D^{*} \right )}{\partial R} \right )} \label{eq:dEdR1} \\
                                                        & = & {\rm Tr}{\left ( \frac{\partial \left (^{2} K \right )}{\partial R} \, ^{2} D^{*} \right )} . \label{eq:dEdR2}
\end{eqnarray}
The second term in Eq.~(\ref{eq:dEdR1}) vanishes because the energy $E^{*}$ has been minimized with respect to all variations about the optimal 2-RDM $^{2} D^{*}$.  Hence, the derivative of the energy $E^{*}$ with respect to $R$, given by Eq.~(\ref{eq:dEdR2}), depends only upon the derivative of the two-electron reduced Hamiltonian matrix with respect to $R$ and the 2-RDM $^{2} D^{*}$.

This result extends the Hellmann-Feynman theorem~\cite{Feynman1939} to approximate $N$-representable sets of 2-RDMs ${\tilde P}^{2}_{N}$.  If ${\tilde P}^{2}_{N} = P^{2}_{N}$, then the result is equivalent to the conventional Hellmann-Feynman theorem.  A similar result was previously presented by Schlimgen and the author~\cite{Schlimgen2018} in the context of computing analytical gradients for variational 2-RDM calculations.  Substituting the elements of the two-electron reduced Hamiltonian matrix for $R$ in Eq.~(\ref{eq:dEdR2}) yields
\begin{equation}
\label{eq:dEdK0}
\frac{\partial E^{*}}{\partial \left (^{2} K^{ij}_{kl} \right )} = {{}^{2} D^{ij}_{kl}}
\end{equation}
or
\begin{equation}
\label{eq:dEdK}
\frac{\partial E^{*}}{\partial \left (^{2} K \right )} = {{}^{2} D}^{*} .
\end{equation}
The response of the minimum energy $E^{*}$ to a variation in an element of the reduced Hamiltonian matrix generates the 2-RDM ${{}^{2} D}^{*}$ in the approximate $N$-representable set ${\tilde P}^{2}_{N}$ that minimizes the energy.

\subsection{Dual Formulation of 2-RDM Theory}

\label{sec:dual}

The minimization of the energy with respect to its 2-RDM in Eq.~(\ref{eq:minE}) can be recast in a dual (or polar) formulation~\cite{Mazziotti2016}
\begin{eqnarray}
{\min_{E,^{2} B_{i}}{E}} &&\label{eq:dual1} \\
{\rm subject~to}~\sum_{i}{  ^{2} B_{i} }&& - \left ( ^{2} K - E \, ^{2} I \right ) \label{eq:dual2}
\end{eqnarray}
where \textcolor{black}{the energy is treated as a variable and the two-particle matrices} $^{2} B_{i}$ provide the $N$-representability conditions that define the set ${\tilde P}^{2}_{N}$
\begin{equation}
\label{eq:B2D2}
{\tilde P}^{2}_{N} = \left \{ ^{2} D~~{\rm such~that}~~{\rm Tr}{\left (^{2} B_{i} \, ^{2} D \right )} \ge 0~~{\rm for~all}~i \right \}.
\end{equation}
The collection of $^{2} B_{i}$ forms a special type of convex set in which  $\alpha \, ^{2} B_{i}$ is a member of the set for all $\alpha \ge 0$, known as a convex cone~\cite{Rockafellar1997}.  Because the cone of $^{2} B_{i}$ determines the approximate convex set of 2-RDMs ${\tilde P}^{2}_{N}$ by Eq.~(\ref{eq:B2D2}), it is said to be the dual (or polar) cone of the set of 2-RDMs~\cite{Mazziotti2012b, Kummer1967} and denoted by $({\tilde P}^{2}_{N})^{*}$.  The dual cone $({\tilde P}^{2}_{N})^{*}$ can represent the 2-positivity conditions~\cite{Coleman1963, Garrod1964},  the 2-positivity plus T1 and T2 conditions~\cite{Erdahl1978, Zhao2004, Mazziotti2005}, or higher-order $N$-representability conditions~\cite{Mazziotti2001c, Mazziotti2006b, Mazziotti2012b}.  In the dual formulation the energy in Eq.~(\ref{eq:dual1}) is minimized subject to fitting the extreme elements $^{2} B_{i}$ of the dual cone to the reduced Hamiltonian $^{2} K$ shifted by the energy $E$~\cite{Mazziotti2016}.  \textcolor{black}{For concreteness the dual-cone matrices $^{2} B_{i}$ of the T2 condition are derived in the Appendix.}

The constraints of the dual formulation can be incorporated into the energy functional through a matrix $^{2} X$ of Lagrange multipliers
\begin{equation}
\label{eq:LM}
E^{*} =  \min_{E,^{2} B_{i}}{ \max_{^{2} X}{ L(E,^{2} B_{i},^{2} X) }} .
\end{equation}
where the Lagrangian $ L(E,^{2} B_{i},^{2} X)$ is
\begin{eqnarray}
 L(E,^{2} B_{i},^{2} X) & = & E - {\rm Tr}{ \left ( ^{2} X  \left ( \sum_{i}{^{2} B_{i}} - ^{2} {\tilde K}  \right )  \right ) }  \\
 ^{2} {\tilde K} & = & ^{2} K - E \, ^{2} I .
\end{eqnarray}
Taking the derivative of the minimum energy $E^{*}$ with respect to the elements of the reduced Hamiltonian matrix yields
\begin{equation}
\label{eq:dEdK2}
\frac{\partial E^{*}}{\partial \left (^{2} K \right ) } = {{}^{2} X^{*}} .
\end{equation}
Comparison of Eq.~(\ref{eq:dEdK2}) with Eq.~((\ref{eq:dEdK}) from the extension of the Hellmann-Feynman theorem reveals a crucial result
\begin{equation}
{{}^{2} X^{*}} = {{}^{2} D^{*}} ,
\end{equation}
{\em namely that the optimal Lagrange multiplier matrix is the 2-RDM.}  The elements of the 2-RDM provide the correct weighting of the constraints to generate the stationary Lagrangian functional for the energy.  \textcolor{black}{Importantly, by Eq.~(\ref{eq:dEdK}) the 2-RDM $^{2} D^{*}$ satisfies the approximate $N$-representability conditions given by the $^{2} B_{i}$ in Eq.~(\ref{eq:B2D2}) without any additional restrictions.}  Hence, while Eqs.~(\ref{eq:dual1}) and~(\ref{eq:dual2}) involve the energy but not the 2-RDM, the 2-RDM can be directly computed from its dual cone through the determination of the Lagrange multipliers.

As shown in Ref.~\cite{Mazziotti2016}, because the interactions in the Hamiltonian scale linearly with the size of the system, the number of $^{2} B_{i}$ matrices from the G2 and T2 conditions scales linearly with the rank $r$ of the one-electron basis set.  \textcolor{black}{The one-body part of the two-particle reduced Hamiltonian is correctly described by only the $^{2} B_{i}$ matrices from the D2 and Q2 conditions, which imply the necessary and sufficient D1 and Q1 conditions~\cite{Mazziotti2012b}, and hence, the $^{2} B_{i}$ matrices from the remaining conditions, G2, T2, and higher $N$-representability conditions,  describe the two-electron Coulomb interaction which scales linearly with system size.}  This is equivalent to a rank reduction since the total number of $^{2} B_{i}$ matrices scales as $r^2$ and $r^3$ for the G and T2 conditions, respectively. This important reduction from the physical scaling of the interaction of electrons reduces the computational cost of the DQGT calculation from $r^9$  to $r^6$.  With the identification of the Lagrange multipliers in Eq.~(\ref{eq:LM}) with the 2-RDM we can use the dual variational 2-RDM theory to compute both the energy and one- and two-electron properties of atoms and molecules at a substantially reduced computational scaling.

\section{Applications}

After a discussion of the methodology, we present applications of the dual variational 2-RDM (v2RDM) method to the hydrogen chain H$_{4}$ and the nitrogen-fixation catalyst FeMoco.

\subsection{Methodology}

The dual v2RDM method is implemented with $^{2} B_{i}$ matrices that correspond to the DQGT conditions with rank reduction as discussed in Ref.~\cite{Mazziotti2016}\textcolor{black}{; $r$ $^{2} B_{i}$ matrices are used for the G2 and T2 conditions.}  The computed 2-RDM allows us to calculate both 1- and 2-electron properties and implement v2RDM-based complete-active-space self-consistent-field (CASSCF) calculations~\cite{Gidofalvi2008, Schlimgen2016, Roos1980}.  In CASSCF a set of molecular orbitals in the valence band, known as active orbitals, is treated by solving the Schr{\"o}dinger equation while the remaining (inactive) orbitals are treated by a mean-field calculation.  Typically, the solution of the Schr{\"o}dinger equation with respect to the space of active orbitals is accomplished by a diagonalization of the Hamiltonian in the basis set of $N$-electron determinants, known as configuration interaction; however, as shown in previous work~\cite{Gidofalvi2008}, configuration interaction can be replaced by a v2RDM method without computation of the wave function.  The v2RDM method must produce the 2-RDM because the active-space 2-RDM is required to perform the orbital rotations of the active and inactive orbitals.  \textcolor{black}{The dual-cone v2RDM method does not depend upon rotations among the active orbitals because the objective and constraints of the optimization problem are invariant to orbital rotations.}  Calculations with only DQG conditions are performed using the boundary-point algorithm in Ref.~\cite{Mazziotti2011} implemented in the Quantum Chemistry Package (QCP) in Maple~\cite{QCP2020}.

\subsection{Results}

\subsubsection{Hydrogen chain}


\begin{figure}
\begin{center}
\includegraphics[scale=.4]{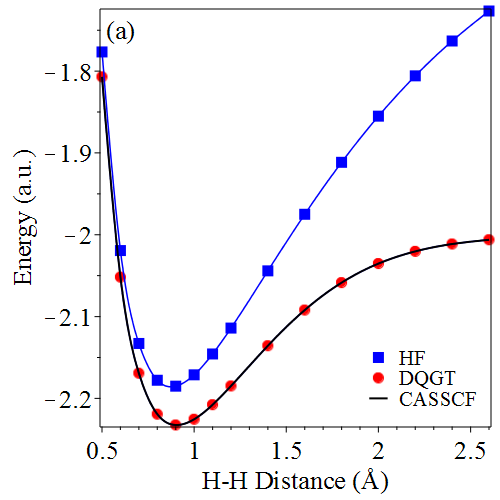}
\includegraphics[scale=.4]{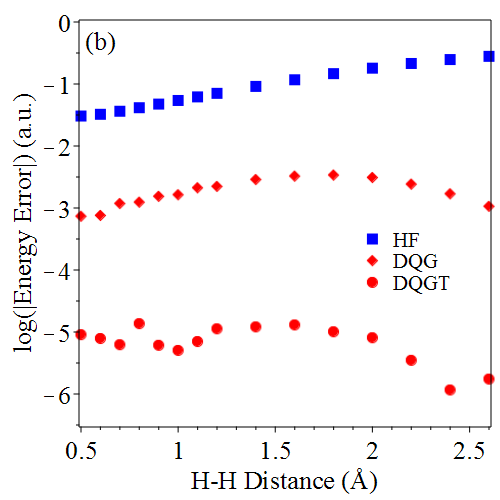}
\end{center}
\caption{The (a) total energy and the (b) energy errors of the potential energy curve of the H$_{4}$ chain are shown for its equally spaced dissociation in a [4,4] active space in the cc-pVQZ basis set.  Errors are relative to CASSCF.  The variational 2-RDM energies with the DQG and DQGT conditions agree with those from CASSCF to about  $10^{-3}$~a.u. and $10^{-5}$~a.u., respectively.}
\label{fig:h4pes}
\end{figure}


\begin{table}[t!]

\caption{For the H$_{4}$ chain the total energy as a function of the distance $R$ between the equally spaced hydrogen atoms is shown with a [4,4] active space in the cc-pVQZ basis set.  Lower-bound energies from the variational 2-RDM method with DQG and DQGT conditions agree with those from the complete-active-space self-consistent-field (CASSCF) method until the third and fifth decimals, respectively.}

\label{t:h4pes}

\begin{ruledtabular}
\begin{tabular}{cccccc}
                                                          &  \multicolumn{4}{c}{Energy (a.u.)}   \\ \cline{2-5}
                                                          &  \multicolumn{2}{c}{Wave Function Methods}  &  \multicolumn{2}{c}{2-RDM Methods} \\ \cline{2-3} \cline{4-5}
   \multicolumn{1}{c}{$R$ (\AA)}  &  \multicolumn{1}{c}{Hartree-Fock}  & \multicolumn{1}{c}{CASSCF} &  \multicolumn{1}{c}{DQG} & \multicolumn{1}{c}{DQGT} \\ \hline
0.8 & -2.177825 &  -2.219294 & -2.220533 & -2.219308 \\
1.0 & -2.171201 &  -2.225488 & -2.227122 & -2.225493 \\
1.2 & -2.114056 &  -2.184800 & -2.187026 & -2.184811 \\
1.6 & -1.974969 &  -2.092035 & -2.095296 & -2.092048 \\
2.0 & -1.855050 &  -2.035218 & -2.038315 & -2.035226 \\
2.4 & -1.763186 &  -2.011306 & -2.012997 & -2.011307 \\
\end{tabular}
\end{ruledtabular}
\end{table}


\begin{table}[t!]

\caption{For the H$_{4}$ chain the metallic character as a function of the distance $R$ between the equally spaced hydrogen atoms is shown with a [4,4] active space in the cc-pVQZ basis set.     The metallic character from the variational 2RDM method with the DQG and DQGT conditions agrees with that from CASSCF to the third and fifth decimals, respectively.  The metallic character is computed from the sum of the squares of the atomic-orbital 1-RDM elements between atoms. }

\label{t:h4mit}

\begin{ruledtabular}
\begin{tabular}{cccccc}
                                                          &  \multicolumn{4}{c}{Metallic Character}   \\ \cline{2-5}
                                                          &  \multicolumn{2}{c}{Wave Function Methods}  &  \multicolumn{2}{c}{2-RDM Methods} \\ \cline{2-3} \cline{4-5}
   \multicolumn{1}{c}{$R$ (\AA)}  &  \multicolumn{1}{c}{Hartree-Fock}  & \multicolumn{1}{c}{CASSCF} &  \multicolumn{1}{c}{DQG} & \multicolumn{1}{c}{DQGT} \\ \hline
0.8 & 0.42226 & 0.33871  &  0.33493  & 0.33872  \\
1.0 & 0.36429 & 0.28100  &  0.27623 &  0.28095  \\
1.2 & 0.36311 & 0.24844  &  0.24214 &  0.24836  \\
1.6 & 0.37279 & 0.15554  &  0.14674 &  0.15550  \\
2.0 & 0.42410 & 0.07323  &  0.07024 &  0.07328  \\
2.4 & 0.49635 & 0.02567  &  0.02700 &  0.02569  \\
\end{tabular}
\end{ruledtabular}
\end{table}


\begin{figure}
\begin{center}
\includegraphics[scale=.4]{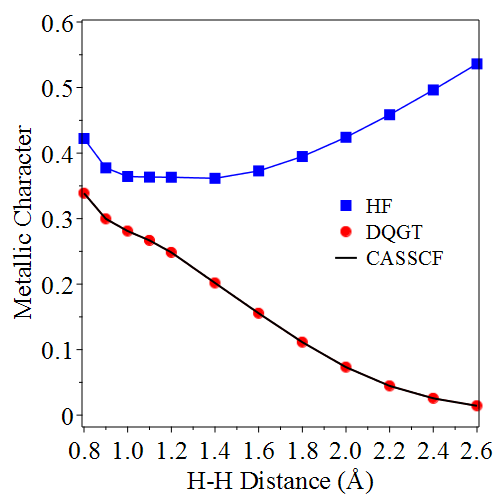}
\end{center}
\caption{The metallic character of the H$_{4}$ chain as a function of the distance $R$ between the equally spaced hydrogen atoms is shown with a [4,4] active space in the cc-pVQZ basis set.    The metallic character is computed from the sum of the squares of the atomic-orbital 1-RDM elements between atoms.  Based on this criterion, the Hartree-Fock theory predicts that the chain remains a metal upon dissociation while the variational 2RDM method with the DQG and DQGT conditions and CASSCF predict a Mott metal-to-insulator transition.}
\label{fig:h4mit}
\end{figure}

Equally spaced metallic hydrogen chains, which exist under high pressure conditions such as the surface of Saturn, undergo a Mott metal-to-insulator transition upon dissociation~\cite{Suhai1994, Sinitskiy2010}.  The transition involves strong electron correlation that is difficult to treat with conventional single-reference wave function methods.  Consequently, hydrogen chains have become a benchmark for treating strongly correlated systems in quantum chemistry and condensed-matter physics.  Here we examine the dissociation of the H$_{4}$ chain in a four-electrons-in-four-orbitals [4,4] active space in the correlation-consistent polarized quadruple-zeta (cc-pVQZ) basis set~\cite{Dunning1989}.  The active space v2RDM method with the DQG and DQGT conditions is compared with the ``exact'' results in this active space from the CASSCF method with the Schr{\"o}dinger equation solved by configuration interaction.  Figure~1 shows the  (a) potential energy curves and the (b) potential-energy-curve errors of the equally spaced H$_{4}$ dissociation.   The lower-bound v2RDM energies with the DQG and DQGT conditions agree with those from CASSCF to about  $10^{-3}$~a.u. and $10^{-5}$~a.u., respectively.

The Mott metal-to-insulator transition in the H$_{4}$ chain can be captured by examining the chain's metallic character as a function of the  distance $R$ between the equally spaced hydrogen atoms.  While various criteria can be selected for metallic character, here we define and compute metallic character from the sum of the squares of the atomic-orbital 1-RDM elements between atoms.   These elements will decay to zero as the metal becomes an insulator.  Based on this criterion as shown in Fig.~\ref{fig:h4mit}, the Hartree-Fock theory predicts that the chain remains a metal upon dissociation while the v2RDM method with the DQG and DQGT conditions and the CASSCF predict a Mott metal-to-insulator transition.  The metallic character data from the v2RDM method with the DQG and DQGT conditions, shown in Table~\ref{t:h4mit}, agree with that from CASSCF to the third and fifth decimals, respectively.


\begin{table}[t!]

\caption{The average charges for each atom type in FeMoco, based on the Mulliken populations computed from the 1-RDM, are shown with the atoms arranged from most positive to most negative.  With the exception of N and O, the electron correlation decreases the absolute values of the charges.}

\label{t:femocochar}

\begin{ruledtabular}
\begin{tabular}{cddd}
                                                     &  \multicolumn{3}{c}{Average Atomic Charge (a.u.)}  \\ \cline{2-4}
    \multicolumn{1}{c}{Atom}  &  \multicolumn{1}{c}{~~~~Hartree-Fock} & \multicolumn{1}{c}{~~~~DQG}  & \multicolumn{1}{c}{~~~~DQGT} \\ \hline
Mo &   1.2450   & 1.1659   &   1.1812 \\
Fe   &   0.5763   & 0.5359   &   0.5446 \\
H    &   0.2022   & 0.1882   &   0.1882  \\
S     & -0.3289   & -0.2961  & -0.3027 \\
O    & -0.5156   & -0.5268  & -0.5267 \\
N    & -0.5455   & -0.5593  & -0.5596 \\
C    & -1.4243   & -1.2843  & -1.2943 \\
\end{tabular}
\end{ruledtabular}
\end{table}


\begin{table}[t!]

\caption{Natural-orbital occupations of the lowest singlet state of FeMoco are reported from the active-space variational 2-RDM method with DQG and DQGT conditions for a [30,30] active space as well as Hartree-Fock (HF) in the DZP basis set.  While many orbitals become slightly less correlated---with occupations closer to 0 and 2---from DQG to DQGT, orbitals 207 through 213 with the exception of 212 become more correlated with the addition of the T2 condition.}

\label{t:femocono}

\begin{ruledtabular}
\begin{tabular}{cccccccc}
      \multicolumn{1}{c}{Orbital}  &  \multicolumn{3}{c}{NO Occupations}  &  \multicolumn{1}{c}{Orbital} & \multicolumn{3}{c}{NO Occupations}  \\ \cline{2-4} \cline{6-8}
    \multicolumn{1}{c}{Index}  &  \multicolumn{1}{c}{HF} & \multicolumn{1}{c}{DQG}  & \multicolumn{1}{c}{DQGT} &
    \multicolumn{1}{c}{Index}  &  \multicolumn{1}{c}{HF} & \multicolumn{1}{c}{DQG}  & \multicolumn{1}{c}{DQGT}  \\ \hline
195 & 2 & 1.9113 & 1.9233 & 210 & 0 & 0.7475 & 0.7782 \\
196 & 2 & 1.8970 & 1.9073 & 211 & 0 & 0.6039 & 0.6278 \\
197 & 2 & 1.8855 & 1.8948 & 212 & 0 & 0.5286 & 0.5255  \\
198 & 2 & 1.8658 & 1.8776 & 213 & 0 & 0.3784 & 0.3886 \\
199 & 2 & 1.8493 & 1.8617 & 214 & 0 & 0.3089 & 0.3062 \\
200 & 2 & 1.8317 & 1.8491 & 215 & 0 & 0.2781 & 0.2801 \\
201 & 2 & 1.8277 & 1.8454 & 216 & 0 & 0.2229 & 0.2084 \\
202 & 2 & 1.8145 & 1.8297 & 217 & 0 & 0.2182 & 0.2062 \\
203 & 2 & 1.7935 & 1.8038 & 218 & 0 & 0.2053 & 0.1886 \\
204 & 2 & 1.7816 & 1.7931 & 219 & 0 & 0.1806 & 0.1670 \\
205 & 2 & 1.7035 & 1.7073 & 220 & 0 & 0.1605 & 0.1442 \\
206 & 2 & 1.6881 & 1.6953 & 221 & 0 & 0.1471 & 0.1371 \\
207 & 2 & 1.5597 & 1.5504 & 222 & 0 & 0.1405 & 0.1273 \\
208 & 2 & 1.2548 & 1.2136 & 223 & 0 & 0.1318 & 0.1187 \\
209 & 2 & 1.0490 & 1.0163 & 224 & 0 & 0.0344 & 0.0275 \\
\end{tabular}
\end{ruledtabular}
\end{table}


\begin{figure}
\begin{center}
\includegraphics[scale=0.8]{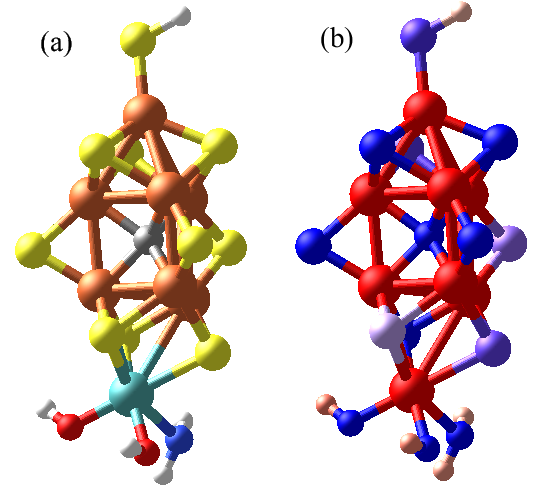}
\end{center}
\caption{The (a) structure of FeMoco and the (b) atomic charges of FeMoco are shown.  The Mulliken charges are computed from the 1-RDM of the variational 2-RDM method with DQGT conditions.  The red indicates positive charge while blue denotes negative charge with the magnitudes of the charges given in Table~\ref{t:femocochar}.}
\label{fig:femocochar}
\end{figure}


\begin{figure}
\begin{center}
\includegraphics[scale=.38]{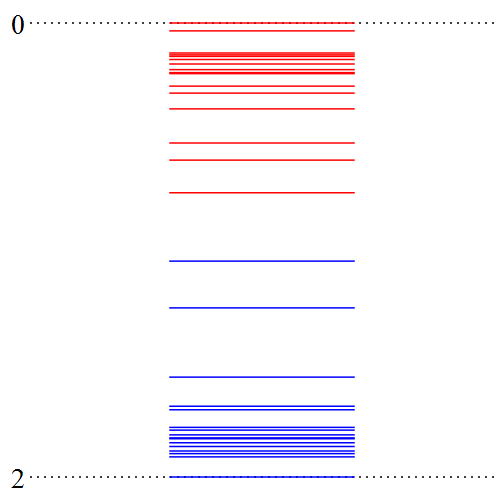}
\end{center}
\caption{The fractional occupations of the natural orbitals of the lowest singlet state of FeMoco are shown visually from the active-space variational 2-RDM method with the DQGT conditions for a [30,30] active space in the DZP basis set.  The blue and red lines correspond to orbitals that are occupied or unoccupied in the Hartree-Fock limit, respectively.}
\label{fig:femocoocc}
\end{figure}


\begin{figure}
\begin{center}
\includegraphics[scale=0.8]{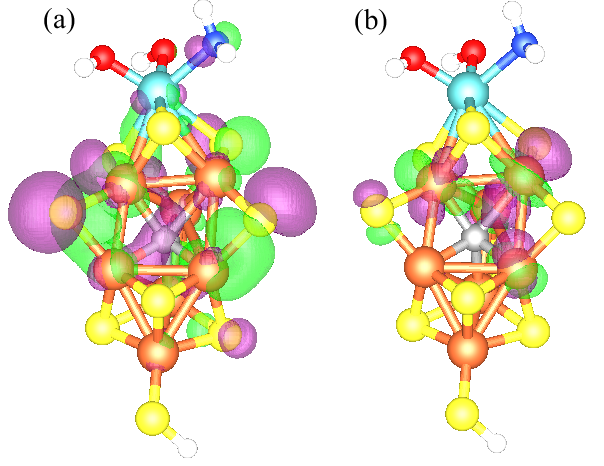}
\end{center}
\caption{The electron density of the 209$^{\rm th}$ natural orbital of FeMoco from the (a) Hartree-Fock method and the (b) variational 2-RDM method with DQGT conditions is displayed.  The highest occupied molecular orbital of the Hartree-Fock method becomes half-filled and much more localized on the Fe centers in the variational 2-RDM method.}
\label{fig:femocono}
\end{figure}

\subsubsection{Nitrogen-fixation catalyst FeMoco}

Nitrogen fixation, the reduction of nitrogen to ammonia, occurs in bacteria by the FeMoco catalyst in the nitrogenase protein~\cite{Lancaster2011, Spatzal2011}.   Despite being mainly treated by density functional theory (DFT), FeMoco is known to be strongly correlated~\cite{Montgomery2018, Stappen2020}.  We recently used the v2RDM method with DQG conditions in a thirty electrons in thirty orbitals [30,30] active space in the polarized double-zeta (DZP) basis set~\cite{Dunning1970} to compute and study the electron correlation in FeMoco~\cite{Montgomery2018}.  The experimental FeMoco was modified by capping the terminal sulfur, nitrogen, and two oxygen with hydrogens, as shown in Fig.~\ref{fig:femocochar}a.   The v2RDM method with DQG conditions uses less than a million variables to represent a wave function with $10^{15}$ degrees of freedom.  Here we extend these results with a v2RDM calculation with the DQGT conditions for a complete-active-space configuration-interaction-like calculation with the optimized orbitals from the DQG calculation.  The addition of the T2 $N$-representability condition to the 2-positivity (DQG) conditions raises the energy from $-17031.7065$~a.u. to $-17031.5855$~a.u., showing the importance of the T2 condition.  Correlation energy from DQGT is $-0.8634$~a.u.  \textcolor{black}{With the computed 2-RDM we can also compute both one- and two-electron properties.  For example, the two-body property $\langle 1/r_{12} \rangle$ is 0.21219~a.u. and 0.21196~a.u. from the DQG and DQGT conditions, respectively.  Both values imply that on average in FeMoco two electrons are approximately 5~a.u. apart.  The electrons are slightly further apart with the DQGT conditions than with the DQG conditions, which is consistent with DQGT exhibiting less electron correlation than DQG.}

Mulliken atomic charges of FeMoco are shown schematically in Fig.~\ref{fig:femocochar}b and numerically in Table~\ref{t:femocochar}.  The Mulliken charges~\cite{Mulliken1955, Reed1985} are computed from the 1-RDM of the v2RDM method with DQG and DQGT conditions as well as the Hartree-Fock method.   Figure~\ref{fig:femocochar}b shows the charges from DQGT conditions with red and blue indicating positive and negative charges, respectively.  We observe that the Mo and Fe atoms are positive while the S and C atoms are negative.  The six-bonded C atom is especially negative with Mulliken charges of $-1.4243$~a.u., $-1.2843$~a.u., and $-1.2943$~a.u. from the Hartree-Fock, DQG, and DQGT methods.   Except for the O and N atoms the electron correlation decreases the absolute values of the atomic charges; in general, the atomic charges from DQGT are slightly larger in magnitude than those from DQG.  The net dipole moment decreases from 2.3799~debyes for the Hartree-Fock method to 1.7495~debyes and 1.6549~debyes for DQG and DQGT conditions, respectively.

Natural-orbital occupations of FeMoco in the [30,30] active space are presented in Fig.~\ref{fig:femocoocc} and Table~\ref{t:femocono}.  Figure~\ref{fig:femocoocc} shows the highly fractional nature of the occupations from v2RDM with DQGT conditions with multiple occupations deviating significantly from 0 and 2.  The blue and red colors, indicating orbitals that are occupied or unoccupied in the Hartree-Fock limit, emphasize that orbitals that are both occupied and unoccupied in the mean-field limit become fractionally occupied.     Table~\ref{t:femocono} reveals that while many orbitals become less correlated---with occupations closer to 0 and 2---from DQG to DQGT, orbitals 207 through 213 with the exception of 212 become more correlated with the addition of the T2 condition.  The von Neumann entropy of the occupation numbers~\cite{Neumann1996}, which is 0 for the Hartree-Fock method, decreases slightly from 5.5689 for DQG to 5.4371 for DQGT.  Figure~\ref{fig:femocono} shows the electron density of the 209$^{\rm th}$ natural orbital from the (a) Hartree-Fock method and the (b) v2RDM method with DQGT conditions.  This orbital is the highest occupied molecular orbital of the Hartree-Fock method, but in the v2RDM method it is half-filled and much more localized on the Fe centers.

\section{Discussion and Conclusions}

The Hellmann-Feynman theorem yields the derivative of a stationary-state energy with respect to an arbitrary parameter without the derivative of the wave function.  Here we examine an analogue of the Hellmann-Feynman theorem for variational 2-RDM theories.  When the energy is variationally minimized with respect to a 2-RDM that is constrained by approximate $N$-representability conditions, the derivative of stationary-state energy with respect to an arbitrary parameter does not depend on the 2-RDM.  The proof relies on the variational principle of the approximate $N$-representable set---specifically, the stationarity of the energy with respect to variations in the 2-RDM constrained by the approximate $N$-representability conditions.  Because the proof is also correct in the limit of the exact $N$-representable set, it can be viewed as a generalization of the traditional Hellmann-Feynman theorem~\cite{Feynman1939}.  Previously, this extension of the Hellmann-Feynman theorem was examined and employed in the context of computing analytical gradients for the v2RDM method.  Here we use the extended Hellmann-Feynman theorem to compute the 2-RDM in the dual v2-RDM theory.

 The v2RDM method has been employed extensively as a polynomially scaling replacement for the configuration interaction solver in CASSCF theory~\cite{Gidofalvi2008, Schlimgen2016, McIsaac2017, Boyn2020, Xie2020, Safaei2018, Sager2020a}.  The computation of the 2-RDM in the dual v2RDM method, presented here, is crucial for its implementation as a solver in the CASSCF theory because at each iteration CASSCF uses the 2-RDM of the active space to perform the self-consistent-field orbital optimization.  While the conventional (primal) formulation of v2RDM with DQGT conditions has a computational scaling of $r^{9}$, the dual formulation of the v2RDM method decreases this scaling to $r^{6}$.  We illustrate the dual active-space v2RDM method in the calculations of both the potential energy surface and the Mott metal-to-insulator transition of a hydrogen chain.  The reduction in computational cost of the v2RDM method in its dual formulation arises from the fact that because the interaction of the Hamiltonian scales linearly with system size, the number of $N$-representability conditions from the G and T2 matrices required to fit this interaction scales linearly with the rank $r$ of the orbital basis set~\cite{Mazziotti2016}.  This linear scaling is valid for higher $N$-representability conditions~\cite{Mazziotti2001c, Mazziotti2012b}, and hence, the dual v2RDM method provides a framework for applying these conditions at reduced computational cost.   Recent work on a variation of the v2RDM method has explored applying a linear scaling number of higher $N$-representability conditions in spin systems~\cite{Haim2020}.

In summary, a dual-cone formulation of the variational 2-RDM method substantially reduces the computational cost of implementing the T2 or higher $N$-representability conditions in both floating-point operations and memory storage~\cite{Mazziotti2016}.  The central result of this paper is that we can compute the 2-RDM in the dual v2RDM method from a generalization of the Hellmann-Feynman theorem.  Moreover, in its Lagrangian formulation the 2-RDM can be identified as the Lagrange multiplier of the Lagrangian functional. We apply the method to computing the energies and properties of strongly correlated electrons---including atomic charges, electron densities, dipole moments, and orbital occupations---in an illustrative hydrogen chain and the nitrogen-fixation catalyst FeMoco.  While there are improvements in upgrading from DQG to DQGT, the degree of electron correlation does not change appreciably, and hence, the DQGT computations reinforce the qualitative understanding of strong correlation in hydrogen chains~\cite{Sinitskiy2010} and FeMoco~\cite{Montgomery2018} from previous studies with DQG.  The dual variational computation of the 2-RDM with the T2 or higher $N$-representability conditions provides a powerful approach to computing strongly correlated molecules and materials with significant applications throughout chemistry and physics.

\begin{acknowledgments}
The author thanks D. Herschbach, H. Rabitz, and A. Mazziotti for their encouragement, and the National Science Foundation, Department of Energy's Office of Basic Energy Sciences, and the Army Research Office for their generous support.
\end{acknowledgments}

\appendix*

{ \color{black}

\section{Dual-Cone Matrices of the T2 Condition}

For concreteness we explicitly derive the dual-cone matrices $^{2} B_{i}$ of the T2 condition.  The T2 condition can be expressed as
\begin{equation}
{\rm Tr}( {\hat T}_{2}^{(i)} \, ^{2} D) \ge 0 ,   ~~~\forall i
\end{equation}
where
\begin{equation}
{\hat T}_{2}^{(i)} = {\hat C}_{i} {\hat C}_{i}^{\dagger} + {\hat C}_{i}^{\dagger}  {\hat C}_{i}
\end{equation}
in which
\begin{equation}
{\hat C}_{i}  = \sum_{jkl}{c^{(i)}_{jkl} {\hat a}^{\dagger}_{j} {\hat a}^{\dagger}_{k} {\hat a}_{l}} .
\end{equation}
The ${\hat a}^{\dagger}_{j}$ and ${\hat a}_{j}$ are second-quantized operators that create and annihilate a fermion in orbital $j$, respectively.  Rearranging the second-quantized operators, we can express the T2 operators as
\begin{equation}
{\hat T}_{2}^{(i)}  = \sum_{jklm}{ ^{2} B^{jk;lm}_{i} {\hat a}^{\dagger}_{j} {\hat a}^{\dagger}_{k} {\hat a}_{m} {\hat a}_{l} }
\end{equation}
in which
\begin{eqnarray}
\label{eq:B2T2}
^{2} B^{jk;lm}_{i} & = & \frac{ {\hat A}_{jk} {\hat A}_{lm} }{4}  \sum_{p} \left ( c_{pjk} c^{*}_{plm} + 4 c_{lkp} c^{*}_{jpm} \right ) \\
& + & \frac{{\hat A}_{jk} {\hat A}_{lm}}{2(N-1)} \delta^{k}_{m} \sum_{pq}{c_{lpq} c^{*}_{jpq}} .
\end{eqnarray}
The antisymmetrization operator ${\hat A}_{jk}$ antisymmetrizes a tensor over the indices $j$ and $k$ by subtracting the permuted tensor from the original tensor.  Each $B_{i}$ in Eq.~(\ref{eq:B2T2}) is an extreme element of the dual cone, and collectively, they enforce the T2 condition.  The  $^{2} B_{i}$  from other $N$-representability conditions are derivable in an analogous fashion.}

\bibliography{v2RDM}

\end{document}